\documentclass[12pt]{article}

\usepackage{microtype}
\usepackage{graphicx}
\usepackage{subfigure}
\usepackage{booktabs} 
\usepackage{url}

\usepackage{hyperref}
\usepackage{breakurl}

\usepackage[accepted]{icml2021}

\icmltitlerunning{Liability and Insurance for Catastrophic Losses: the Nuclear Power Precedent and Lessons for AI}

\begin{document}
\thispagestyle{plain}
\pagestyle{plain}

\icmltitle{Liability and Insurance for Catastrophic Losses: the Nuclear Power Precedent and Lessons for AI}



\icmlsetsymbol{equal}{*}

\begin{icmlauthorlist}
\icmlauthor{Cristian Trout}{ind}
\end{icmlauthorlist}

\icmlaffiliation{ind}{Independent Researcher, Cambridge Boston Alignment Initiative}

\icmlcorrespondingauthor{Cristian Trout}{ctroutcsi@gmail.com}

\icmlkeywords{AI Policy, Liability, Insurance, ICML}




\printAffiliationsAndNotice{} 

\begin{abstract}
\textit{As AI systems become more autonomous and capable, experts warn of them potentially causing catastrophic losses. Drawing on the successful precedent set by the nuclear power industry, this paper argues that developers of frontier AI models should be assigned limited, strict, and exclusive third party liability for harms resulting from Critical AI Occurrences (CAIOs) – events that cause or easily could have caused catastrophic losses. Mandatory insurance for CAIO liability is recommended to overcome developers' judgment-proofness, mitigate winner's curse dynamics, and leverage insurers' quasi-regulatory abilities. Based on theoretical arguments and observations from the analogous nuclear power context, insurers are expected to engage in a mix of causal risk-modeling, monitoring, lobbying for stricter regulation, and providing loss prevention guidance in the context of insuring against heavy-tail risks from AI. While not a substitute for regulation, clear liability assignment and mandatory insurance can help efficiently allocate resources to risk-modeling and safe design, facilitating future regulatory efforts.}
\end{abstract}

\section{Background}

With generative AI's sudden deployment across society, there has been a sharp rise in AI related incidents \cite{oecd_oecd_2024}. Many experts fear that more advanced AI could cause catastrophic losses \cite{grace_thousands_2024}. Agentic AI is arguably of greatest concern \cite{carlsmith_is_2022,chan_harms_2023}, the sparks of which can arguably be seen in today's generative large language models.

With the exception of Weil \citeyearpar{weil_tort_2024} which addresses existential risks, previous research on liability regimes for AI has largely ignored the possibility of catastrophic losses. This paper seeks to fill this gap, focusing on catastrophic yet still insurable risks. I reply to Weil and extend my analysis to strictly uninsurable risks elsewhere \citep{trout_insuring_2024}.

Research on non-catastrophic losses provides a useful baseline. It's been argued that as models become more autonomous, applying \textit{respondeat superior}, a form of strict vicarious liability, is appropriate \cite{lior_ai_2019}. Strict liability has also been recommended under Fletcher's framework of non-reciprocal harms: parties are to be held strictly liability when they generate ``disproportionate, excessive risk of harm, relative to the victim's risk-creating activity" \citeyearpar[96]{lior_ai_2020}. It's been suggested that AI developers (e.g. OpenAI) or manufacturers (e.g. Tesla) should be the liable parties, given they're better able to absorb or spread costs \citep[146-147]{vladeck_machines_2014}.

Under welfare-maximization analyses, researchers find that developers tend to be the least-cost avoiders \cite{buiten_eu_2021, buiten_law_2023, price_ii_locating_2023}. While operators can play a role in mitigating harms, it’s found to decrease as model autonomy increases or the information asymmetry between developer and operator increases – precisely the case we’re concerned with here. This suggests developers should be assigned greater liability. Concerns of slower adoption of safer technologies, or a possible chilling effect on innovation moderate this suggestion, but scholars express uncertainty on the matter \citep[6-7, 9]{buiten_law_2023}\cite{maliha_artificial_2021}. Strict liability is also often recommended for its greater legal predictability and reduction in litigation (e.g. \citep[147]{vladeck_machines_2014}). Liability would create new barriers to entry, but scholars have noted this industry will produce natural monopolies regardless \cite{lior_ai_2020,vipra_market_2023}; consumer interests are arguably best advanced through stricter regulation and liability \cite{narechania_antimonopoly_2024}.

\section{Liability: Strict, Exclusive, and Limited}

Drawing on the precedent set by the nuclear power industry, this paper argues developers of frontier foundation models should be assigned limited, strict, and exclusive third party liability for ``Critical AI Occurrences" (CAIOs). Criteria for CAIOs might include: monetary thresholds, specific technical failure modes, threat modalities (e.g. CBRN), and persons/objects affected (e.g. critical infrastructure).

Given the baseline set by non-catastrophic losses and the importance of robustly internalizing catastrophic negative externalities, strict liability is a natural choice. It's claimed that developers are even more clearly the least-cost avoiders, and, should such risks materialize, these systems will certainly have generated starkly ``disproportionate, excessive risk of harm, relative to the victim's risk-creating activity" \citep[96]{lior_ai_2020}. ``Strict liability" refers here to regimes under which most defenses relating to claimant conduct or developer fault are waived (defenses based on failure to mitigate damages or wrongful causation by the claimant are not waived).

Capping liability and making it exclusive helps ensure insurability. A cap might appear to generate moral hazard, but \textit{de facto}, liability is typically limited by the solvency of the liable party \citep[355]{meehan_lessons_nodate}. One thus improves over this by setting a cap well above the expected solvency of liable parties and mandating carrying insurance \citep{logue_solving_1993}. Furthermore, given the uncertainty and size of the risks in question, premiums would be prohibitive without a cap, running counter to the policy goal of encouraging responsible innovation. Additionally, safety investment for heavy-tail risks is likely inelastic; a cap much greater than e.g. \$200B is unlikely to yield greater care taken, but simply lower activity levels. Finally, frontier AI's potential positive externalities justify keeping premiums affordable. Exclusive liability further reduces litigation and insurers' aggregate risk: this keeps premiums down while maximizing capacity insurers can commit to the industry. These elements (and the reasoning behind them) mirror the nuclear power precedent \citep[sec. 2.3.8]{united_states_nuclear_regulatory_commission_price-anderson_2021}.

\section{Mandatory Insurance}

Mandating developers carry CAIO insurance is recommended for several reasons. First, as noted above, developers are judgment-proof to an extent: CAIO liability alone will fail to sufficiently incentivize caution. Second, the most successful developers are likely those that most underestimate the risks (a form of the winner’s curse \citep{van_der_merwe_tort_2024}). This underestimation is likely substantial given how large and uncertain the risks are. By turning very uncertain \textit{ex post }costs into very certain \textit{ex ante }costs, insurance mitigates this perverse selection effect. (It's been observed that insurers charge a premium for uncertainty, a heuristic likely aimed at avoiding their own winner’s curse \citep{mumpower_risk_1991}). Third, mandating insurance eliminates adverse selection and further spreads risk, thus keeping premiums down (minimizing barriers to entry). Fourth, insurers are likely to play a significant quasi-regulatory role (possibly more effectively than public institutions presently could), all while providing a testbed for future regulations.

This last bears elaborating. While some scholars tout the virtues of ``outsourcing regulation" to (commercial) insurers \citep{ben-shahar_outsourcing_2012,baker_regulation_2012}, others note limits \citep{abraham_limits_2022}. Insurers' standard tools – risk-based pricing, underwriting, monitoring, \textit{ex post} loss control – \textit{mitigate} but don't \textit{eliminate} moral hazard. For a variety of reasons (e.g. the cost of monitoring, the limits of actuarial risk-modeling, the cost of causal risk-modeling), insurers will fail to incent socially optimal levels of care with these tools alone.

However, insurers also often have incentive to reduce aggregate risk in a predictable manner and directly control insured’s behavior \citep[III]{abraham_limits_2022}. Hence we find them funding safety research, lobbying for stricter regulation, counseling on loss prevention, and enforcing private safety codes. When successful, these measures can improve loss prevention over the baseline (liability without insurance). Unfortunately obstacles often impede such efforts: safety research and stricter regulation are public goods, requiring competing insurers to coordinate; private advice and safety codes can be appropriated by insureds and competitors, cause policyholder backlash, and expose insurers to liability.

This paper claims none of these obstacles are present in the context of insuring against heavy tail risks, and we should expect a mix of causal risk-modeling, monitoring, safety research, lobbying for stricter regulation, and private safety guidance from insurers. Causal risk-modeling because actuarial data will be insufficient for such rare events (cf. causal risk-modeling in nuclear insurance underwriting and premium pricing \citep{mustafa_insuring_2017}\citep[ch. 4 sec. VII]{gudgel_insurance_2022}). Monitoring, again due to a lack of actuarial data and the need to reduce information asymmetries (cf. regular inspections by nuclear insurers with specialized engineers \citeyearpar[ch. 4 sec. VI.C]{gudgel_insurance_2022}). Safety research and lobbying regulators because insurers will almost certainly have to pool their capacity in order to offer coverage, eliminating competition and with it, coordination problems (cf. American Nuclear Insurer’s (ANI) monopoly on third party liability \citeyearpar[ch. 4 sec. VII.A]{gudgel_insurance_2022}). Loss prevention guidance, because it can’t be appropriated or drive away customers here: there will be little competition and the insurance is mandatory (cf. ANI sharing inspection reports and recommendations with policyholders \citeyearpar[ch. 4 sec. VII.A.2]{gudgel_insurance_2022}).

Concerns of monopolistic practices can be mitigated through premium regulation or encouraging self-insurance through a mutual. (That \textit{mutuals} successfully play this quasi-regulatory role is less contentious \citep[sec. I.B.3]{abraham_limits_2022}). For example, the nuclear utilities mutual Nuclear Electric Insurance Limited (NEIL) competes with ANI for certain lines of coverage \citeyearpar[147-148]{gudgel_insurance_2022}. NEIL members are required to join the Institute of Nuclear Power Operations (INPO), a body through which the industry self-regulates \citeyearpar[ch. 4 sec. VI.C]{gudgel_insurance_2022}. Like ANI, INPO conducts inspections, as well as collects and shares best practices. Perhaps surprisingly, commentators have found the INPO to be very effective \citeyearpar[ch. 4 sec. VII.B]{gudgel_insurance_2022}\citep{rees_hostages_1996} and to wield considerable power (e.g. orchestrating a COO and CEO’s firing by exceptionally sharing unheeded warnings with the utility’s board and the Nuclear Regulatory Commission \citeyearpar[110-118]{rees_hostages_1996}). Besides encouraging mutualization of risk, the other features that successfully enabled self-regulation in the nuclear industry (peer pressure, shared liability in the form of fragile public trust, and a perennial threat of regulatory action) could be stressed for the frontier AI industry \citep{gunningham_industry_1997}.

Thus, in the context of insuring against catastrophic risks from AI, we can expect regulation via insurance to be effective – possibly more effective than what public institutions are currently capable of. (Certainly insurers' involvement is expected to complement efforts by public institutions). Government agencies are not disciplined by the market or a profit motive, and are comparatively under-resourced. Courts are purely \textit{ex post}, and reliant on the costly, ad hoc litigation process \citep[198-199]{ben-shahar_outsourcing_2012}.

\section{Conclusion}

Advances in AI appear to pose risks as large and uncertain as those that nuclear power did (and largely still does) when it was first introduced. This paper claims we can learn from the successful precedent set by that industry's liability and insurance regime. It argues that in similarly heavy-tail risk contexts, insurers are expected to play a socially beneficial quasi-regulatory role. It's \textit{not} claimed this can \textit{fully substitute} for regulation. Rather, it's claimed that by assigning clear liability and mandating insurance for these emerging risks, lawmakers can leverage the market’s strength for aggregating information, and efficiently allocating resources to model risks and develop safe design. Regulating efficiently is easier when risks are clearer and safe design is available. Government reinsurance or indemnification schemes (as seen in, respectively, the terrorism risk \citep{federal_insurance_office_report_2022} and nuclear power contexts \citep[sec. 3.2]{united_states_nuclear_regulatory_commission_price-anderson_2021}) are envisioned to handle strictly uninsurable risks: I address these elsewhere \citep{trout_insuring_2024}.

\section*{Acknowledgements}
This work was supported by the Cambridge Boston Alignment Initiative. Thanks to Thomas Larsen, and Mackenzie Arnold for helpful guidance early on. Thanks to Mackenzie Arnold, Gabriel Weil, Trevor Levin, Siti Liyana Azman, and Ketan Ramakrishnan for helpful feedback.


\bibliography{references}

\begin{thebibliography}{28}
\providecommand{\natexlab}[1]{#1}
\providecommand{\url}[1]{\texttt{#1}}
\expandafter\ifx\csname urlstyle\endcsname\relax
  \providecommand{\doi}[1]{doi: #1}\else
  \providecommand{\doi}{doi: \begingroup \urlstyle{rm}\Url}\fi

\bibitem[Abraham \& Schwarcz(2022)Abraham and Schwarcz]{abraham_limits_2022}
Abraham, K.~S. and Schwarcz, D.
\newblock The {Limits} of {Regulation} by {Insurance}.
\newblock \emph{Indiana Law Journal}, 98:\penalty0 215, 2022.
\newblock URL \url{https://heinonline.org/HOL/Page?handle=hein.journals/indana98&id=228&div=&collection=}.

\bibitem[Baker \& Swedloff(2012)Baker and Swedloff]{baker_regulation_2012}
Baker, T. and Swedloff, R.
\newblock Regulation by {Liability} {Insurance}: {From} {Auto} to {Lawyers} {Professional} {Liability}.
\newblock \emph{UCLA Law Review}, 60:\penalty0 1412, 2012.
\newblock URL \url{https://heinonline.org/HOL/Page?handle=hein.journals/uclalr60&id=1454&div=&collection=}.

\bibitem[Ben-Shahar \& Logue(2012)Ben-Shahar and Logue]{ben-shahar_outsourcing_2012}
Ben-Shahar, O. and Logue, K.~D.
\newblock Outsourcing {Regulation}: {How} {Insurance} {Reduces} {Moral} {Hazard}.
\newblock \emph{Michigan Law Review}, 111\penalty0 (2):\penalty0 197--248, 2012.
\newblock ISSN 0026-2234.
\newblock URL \url{https://www.jstor.org/stable/41703440}.
\newblock Publisher: Michigan Law Review Association.

\bibitem[Buiten et~al.(2021)Buiten, de~Streel, and Peitz]{buiten_eu_2021}
Buiten, M., de~Streel, A., and Peitz, M.
\newblock {EU} liability rules for the age of {Artificial} {Intelligence}, 2021.
\newblock URL \url{https://cerre.eu/publications/eu-liability-rules-age-of-artificial-intelligence-ai/}.

\bibitem[Buiten et~al.(2023)Buiten, de~Streel, and Peitz]{buiten_law_2023}
Buiten, M., de~Streel, A., and Peitz, M.
\newblock The law and economics of {AI} liability.
\newblock \emph{Computer Law \& Security Review}, 48:\penalty0 105794, April 2023.
\newblock ISSN 0267-3649.
\newblock \doi{10.1016/j.clsr.2023.105794}.
\newblock URL \url{https://www.sciencedirect.com/science/article/pii/S0267364923000055}.

\bibitem[Carlsmith(2022)]{carlsmith_is_2022}
Carlsmith, J.
\newblock Is {Power}-{Seeking} {AI} an {Existential} {Risk}?, June 2022.
\newblock URL \url{http://arxiv.org/abs/2206.13353}.
\newblock arXiv:2206.13353 [cs].

\bibitem[Chan et~al.(2023)Chan, Salganik, Markelius, Pang, Rajkumar, Krasheninnikov, Langosco, He, Duan, Carroll, Lin, Mayhew, Collins, Molamohammadi, Burden, Zhao, Rismani, Voudouris, Bhatt, Weller, Krueger, and Maharaj]{chan_harms_2023}
Chan, A., Salganik, R., Markelius, A., Pang, C., Rajkumar, N., Krasheninnikov, D., Langosco, L., He, Z., Duan, Y., Carroll, M., Lin, M., Mayhew, A., Collins, K., Molamohammadi, M., Burden, J., Zhao, W., Rismani, S., Voudouris, K., Bhatt, U., Weller, A., Krueger, D., and Maharaj, T.
\newblock Harms from {Increasingly} {Agentic} {Algorithmic} {Systems}.
\newblock In \emph{Proceedings of the 2023 {ACM} {Conference} on {Fairness}, {Accountability}, and {Transparency}}, {FAccT} '23, pp.\  651--666, New York, NY, USA, June 2023. Association for Computing Machinery.
\newblock ISBN 9798400701924.
\newblock \doi{10.1145/3593013.3594033}.
\newblock URL \url{https://dl.acm.org/doi/10.1145/3593013.3594033}.

\bibitem[Commission(2021)]{united_states_nuclear_regulatory_commission_price-anderson_2021}
Commission, U. S. N.~R.
\newblock The {Price}-{Anderson} {Act}: 2021 {Report} to {Congress}, {Public} {Liability} {Insurance} and {Indemnity} {Requirements} for an {Evolving} {Commercial} {Nuclear} {Industry}.
\newblock 2021.
\newblock URL \url{https://www.nrc.gov/docs/ML2133/ML21335A064.pdf}.

\bibitem[Federal Insurance~Office(2022)]{federal_insurance_office_report_2022}
Federal Insurance~Office, U. D. o. t.~T.
\newblock Report on the {Effectiveness} of the {Terrorism} {Risk} {Insurance} {Program}.
\newblock Technical report, June 2022.
\newblock URL \url{https://home.treasury.gov/system/files/311/2022%20Program%20Effectiveness%20Report%20%28FINAL%29.pdf}.

\bibitem[Grace et~al.(2024)Grace, Stewart, Sandkühler, Thomas, Weinstein-Raun, and Brauner]{grace_thousands_2024}
Grace, K., Stewart, H., Sandkühler, J.~F., Thomas, S., Weinstein-Raun, B., and Brauner, J.
\newblock Thousands of {AI} {Authors} on the {Future} of {AI}, April 2024.
\newblock URL \url{http://arxiv.org/abs/2401.02843}.
\newblock arXiv:2401.02843 [cs].

\bibitem[Gudgel(2022)]{gudgel_insurance_2022}
Gudgel, J.~E.
\newblock Insurance as a {Private} {Sector} {Regulator} and {Promoter} of {Security} and {Safety}: {Case} {Studies} in {Governing} {Emerging} {Technological} {Risk} {From} {Commercial} {Nuclear} {Power} to {Health} {Care} {Sector} {Cybersecurity}.
\newblock 2022.
\newblock URL \url{https://hdl.handle.net/1920/13083}.

\bibitem[Gunningham \& Rees(1997)Gunningham and Rees]{gunningham_industry_1997}
Gunningham, N. and Rees, J.
\newblock Industry {Self}-{Regulation}: {An} {Institutional} {Perspective}.
\newblock \emph{Law \& Policy}, 19\penalty0 (4):\penalty0 363--414, 1997.
\newblock ISSN 1467-9930.
\newblock \doi{10.1111/1467-9930.t01-1-00033}.
\newblock URL \url{https://onlinelibrary.wiley.com/doi/abs/10.1111/1467-9930.t01-1-00033}.
\newblock \_eprint: https://onlinelibrary.wiley.com/doi/pdf/10.1111/1467-9930.t01-1-00033.

\bibitem[Lior(2019)]{lior_ai_2019}
Lior, A.
\newblock {AI} {Entities} as {AI} {Agents}: {Artificial} {Intelligence} {Liability} and the {AI} {Respondeat} {Superior} {Analogy}.
\newblock \emph{Mitchell Hamline Law Review}, 46:\penalty0 1043, 2019.
\newblock URL \url{https://heinonline.org/HOL/Page?handle=hein.journals/wmitch46&id=1043&div=&collection=}.

\bibitem[Lior(2020)]{lior_ai_2020}
Lior, A.
\newblock {AI} {Strict} {Liability} vis-a-vis {AI} {Monopolization}.
\newblock \emph{Columbia Science and Technology Law Review}, 22:\penalty0 90, 2020.
\newblock URL \url{https://heinonline.org/HOL/Page?handle=hein.journals/cstlr22&id=90&div=&collection=}.

\bibitem[Logue(1993)]{logue_solving_1993}
Logue, K.~D.
\newblock Solving the {Judgment}-{Proof} {Problem}.
\newblock \emph{Texas Law Review}, 72:\penalty0 1375, 1993.
\newblock URL \url{https://heinonline.org/HOL/Page?handle=hein.journals/tlr72&id=1409&div=&collection=}.

\bibitem[Maliha et~al.(2021)Maliha, Gerke, Cohen, and Parikh]{maliha_artificial_2021}
Maliha, G., Gerke, S., Cohen, I.~G., and Parikh, R.~B.
\newblock Artificial {Intelligence} and {Liability} in {Medicine}: {Balancing} {Safety} and {Innovation}.
\newblock \emph{The Milbank Quarterly}, 99\penalty0 (3):\penalty0 629--647, September 2021.
\newblock ISSN 0887-378X.
\newblock \doi{10.1111/1468-0009.12504}.
\newblock URL \url{https://www.ncbi.nlm.nih.gov/pmc/articles/PMC8452365/}.

\bibitem[Meehan()]{meehan_lessons_nodate}
Meehan, T.
\newblock Lessons {From} the {Price}-{Anderson} {Nuclear} {Industry} {Indemnity} {Act} for {Future} {Clean} {Energy} {Compensatory} {Models}.

\bibitem[Mumpower(1991)]{mumpower_risk_1991}
Mumpower, J.~L.
\newblock Risk, {Ambiguity}, {Insurance}, and the {Winner}'s {Curse}.
\newblock \emph{Risk Analysis}, 11\penalty0 (3):\penalty0 519--522, 1991.
\newblock ISSN 1539-6924.
\newblock \doi{10.1111/j.1539-6924.1991.tb00637.x}.
\newblock URL \url{https://onlinelibrary.wiley.com/doi/abs/10.1111/j.1539-6924.1991.tb00637.x}.
\newblock \_eprint: https://onlinelibrary.wiley.com/doi/pdf/10.1111/j.1539-6924.1991.tb00637.x.

\bibitem[Mustafa(2017)]{mustafa_insuring_2017}
Mustafa, R.
\newblock \emph{Insuring nuclear risk}.
\newblock PhD thesis, July 2017.
\newblock Available from INIS: http://inis.iaea.org/search/search.aspx?orig\_q=RN:50083099 INIS Reference Number: 50083099.

\bibitem[Narechania \& Sitaraman(2024)Narechania and Sitaraman]{narechania_antimonopoly_2024}
Narechania, T.~N. and Sitaraman, G.
\newblock An {Antimonopoly} {Approach} to {Governing} {Artificial} {Intelligence}, January 2024.
\newblock URL \url{https://papers.ssrn.com/abstract=4755377}.

\bibitem[OECD(2024)]{oecd_oecd_2024}
OECD.
\newblock {OECD} {AI} {Incidents} {Monitor} ({AIM}), 2024.
\newblock URL \url{https://oecd.ai/en/incidents}.

\bibitem[Price~II \& Cohen(2023)Price~II and Cohen]{price_ii_locating_2023}
Price~II, W.~N. and Cohen, I.~G.
\newblock Locating {Liability} for {Medical} {AI}, July 2023.
\newblock URL \url{https://papers.ssrn.com/abstract=4517740}.

\bibitem[Rees(1996)]{rees_hostages_1996}
Rees, J.~V.
\newblock \emph{Hostages of {Each} {Other}: {The} {Transformation} of {Nuclear} {Safety} since {Three} {Mile} {Island}}.
\newblock University of Chicago Press, Chicago, IL, June 1996.
\newblock ISBN 978-0-226-70688-7.
\newblock URL \url{https://press.uchicago.edu/ucp/books/book/chicago/H/bo3618989.html}.

\bibitem[Trout(2024)]{trout_insuring_2024}
Trout, C.
\newblock Insuring {Uninsurable} {Risks} from {AI}: {Government} as {Insurer} of {Last} {Resort}.
\newblock In \emph{Generative {AI} and {Law} {Workshop} at the {International} {Conference} on {Machine} {Learning}}, Vienna, Austria, 2024.

\bibitem[van~der Merwe et~al.(2024)van~der Merwe, Ramakrishnan, and Anderljung]{van_der_merwe_tort_2024}
van~der Merwe, M., Ramakrishnan, K., and Anderljung, M.
\newblock Tort {Law} and {Frontier} {AI} {Governance}, 2024.
\newblock URL \url{https://www.lawfaremedia.org/article/tort-law-and-frontier-ai-governance}.

\bibitem[Vipra \& Korinek(2023)Vipra and Korinek]{vipra_market_2023}
Vipra, J. and Korinek, A.
\newblock Market {Concentration} {Implications} of {Foundation} {Models}: {The} {Invisible} {Hand} of {ChatGPT}.
\newblock 2023.

\bibitem[Vladeck(2014)]{vladeck_machines_2014}
Vladeck, D.~C.
\newblock Machines without {Principals}: {Liability} {Rules} and {Artificial} {Intelligence}.
\newblock \emph{Washington Law Review}, 89:\penalty0 117, 2014.
\newblock URL \url{https://heinonline.org/HOL/Page?handle=hein.journals/washlr89&id=124&div=&collection=}.

\bibitem[Weil(2024)]{weil_tort_2024}
Weil, G.
\newblock Tort {Law} as a {Tool} for {Mitigating} {Catastrophic} {Risk} from {Artificial} {Intelligence}, January 2024.
\newblock URL \url{https://papers.ssrn.com/abstract=4694006}.

\end{thebibliography}
\bibliographystyle{icml2021}

\end{document}